# Polarized electron-beam acceleration driven by vortex laser pulses

Yitong Wu[1,2], Liangliang Ji[1,3], Xuesong Geng[1], Qin Yu[1], Nengwen Wang[1], Bo Feng[1], Zhao Guo[1], Weiqing Wang[1], Chengyu Qin[1], Xue Yan[1], Lingang Zhang[1], Johannes Thomas[5], Anna Hützen[6,7], Markus Büscher[6,7], Peter Rakitzis[8,9], Alexander Pukhov[5], Baifei Shen,[1,3,4] and Ruxin Li[1,3,10]

[1]*State Key Laboratory of High Field Laser Physics, Shanghai Institute of Optics and Fine Mechanics, Chinese Academy of Sciences, Shanghai 201800, China*

[2]*Center of Materials Science and Optoelectronics Engineering, University of Chinese Academy of Sciences, Beijing 100049, China*

[3]*CAS Center for Excellence in Ultra-intense Laser Science, Shanghai 201800, China*

[4]*Shanghai Normal University, Shanghai 200234, China*

[5]*Institut fuer Theoretische Physik I, Heinrich-Heine-Universität Düsseldorf, 40225 Düsseldorf, Germany*

[6]*Peter Grünberg Institut (PGI-6), Forschungszentrum Jülich, Wilhelm-Johnen-Str. 1, 52425 Jülich, Germany*

[7]*Institut für Laser- und Plasmaphysik, Heinrich-Heine-Universität Düsseldorf, 40225 Düsseldorf, Germany*

[8]*Department of Physics, University of Crete, 71003 Heraklion-Crete, Greece*

[9]*Institute of Electronic Structure and Laser, Foundation for Research and Technology-Hellas, 71110, Heraklion-Crete, Greece*

[10]*Shanghai Tech University, Shanghai 201210, China*

## Abstract

We propose a new approach based on an all-optical set-up for generating relativistic polarized electron beams via vortex Laguerre-Gaussian (LG) laser-driven wakefield acceleration. Using a pre-polarized gas target, we find that the topology of the vortex wakefield resolves the depolarization issue of the injected electrons. In full three-dimensional particle-in-cell simulations, incorporating the spin dynamics via the Thomas-Bargmann Michel Telegdi equation, the LG laser preserves the electron spin polarization by more than 80% while assuring efficient electron injection. The method releases the limit on beam flux for polarized electron acceleration and promises more than an order of magnitude boost in peak flux, as compared to Gaussian beams. These results suggest a promising table-top method to produce energetic polarized electron beams.

E-mail: jill@siom.ac.cn, bfshen@ mail.shcnc.ac.cn and ruxinli@mail.shcnc.ac.cn

# 1. Introduction

Spin is an intrinsic form of angular momentum carried by elementary particles [1]. Numerous studies in particle physics and material science have been carried out using spin-polarized electron beams [2-6]. Generally, generating polarized electrons requires conventional accelerators (Storage ring or Linac) that are typically very large in scale and budget [2,7,8]. In some cases, it also needs sufficient long time to attain high polarization (couple of hours for storage rings [9]). Thanks to the rapid development of laser technology, the focal light intensities are now well beyond $10^{20}$W/cm$^3$ [10-12], which paves a new path to obtain high energy electron beams based on the concept of laser-driven wakefield acceleration(LWFA) [13-15]. The latter, due to the extremely high acceleration gradient, promises a more compact and cost-efficient approach for electron acceleration. However, for the realization of a laser-driven accelerator for polarized electron beams several challenges need to be addressed: i) Since a significant build-up of electron polarization from an initially unpolarized target during laser acceleration does not happen [16,17], it requires the use of a gas target where the electron spins are already aligned before laser irradiation. ii) Polarization losses during the injection of as many as possible electrons into a bubble structure and iii) subsequent acceleration in the wake field must be kept under control and minimized.

With regard to polarized electron targets suitable for LWFA we have witnessed promising developments in recent years. Electron spin polarization in strong-field ionization of atoms has been widely studied both theoretically [18-23] and experimentally [18-20]. Gas targets with spin polarization ~40% have been achieved using near-infrared light [18-20]. It is also predicted that the polarization can reach 90% by adapting ultraviolet (UV) light [19,20]. The most promising approach seems to be a technique employing the UV photodissociation of hydrogen halides [16,24,25], reaching gas densities of approx. $10^{19}$cm$^{-3}$ at high polarization of the electrons in the hydrogen and halogen atoms [26-28](See detail in Section 2).

The perspective of all-optical laser-driven polarized electron acceleration therefore relies on addressing the key issues ii) and iii), i.e., electron depolarization during LWFA. As shown below, spin depolarization mainly happens in the injection phase, which is in line with previous studies [29]. Recently, a method is proposed to mitigate this issue by fine tuning the focal position of the Gaussian laser beam to weaken the electron injection [30]. Here we propose that compared to Gaussian lasers, vortex laser beams are capable of creating a unique topology in LWFA that significantly suppresses the beam depolarization without sacrifice of the injected electron charge. Based on the proposed all-optical experiment set-up available for present techniques, we demonstrate in full three-dimensional particle-in-cell simulations the generation of high polarization electrons at very high beam charge. Such laser beams are readily accessible, for example, with the Laguerre Gaussian (LG) mode [31,32].

The paper is organized as follows: In Section 2 we give descriptions of the proposal to generate fully polarized electron target and parameters setup for simulations. In Section 3 we present the results of simulations and analysis. Finally in Section 4, we give a brief summary.

## 2. Scheme descriptions and Simulation Setup

As stated above, preparation a pre-polarized electron target is crucial to our scheme. Known from previous literatures, the photodissociation of hydrogen halides with circularly polarized UV light yields highly spin-polarized hydrogen and halogen atoms [26,27], at gas densities of at least $10^{19}$ cm$^{-3}$ [28]. The polarization can approach 100% for specific photodissociation wavelengths (for HCl this is near 213 nm), and if the molecular bonds are aligned prior to photodissociation [25]; otherwise, if the bonds are isotropic, the polarization is reduced to 40%. However, the electronic polarization of the halogen atoms is very low; specifically, the core electrons are completely unpolarized. Therefore, to maximize the polarization of all accelerated electrons, the halogen atoms can be ionized with a resonant laser (e.g. via 2+1 resonance-enhanced multiphoton ionization (REMPI), for Cl atoms at 234.62 nm [26]) and removed from the target volume before irradiation by the accelerating laser pulse.

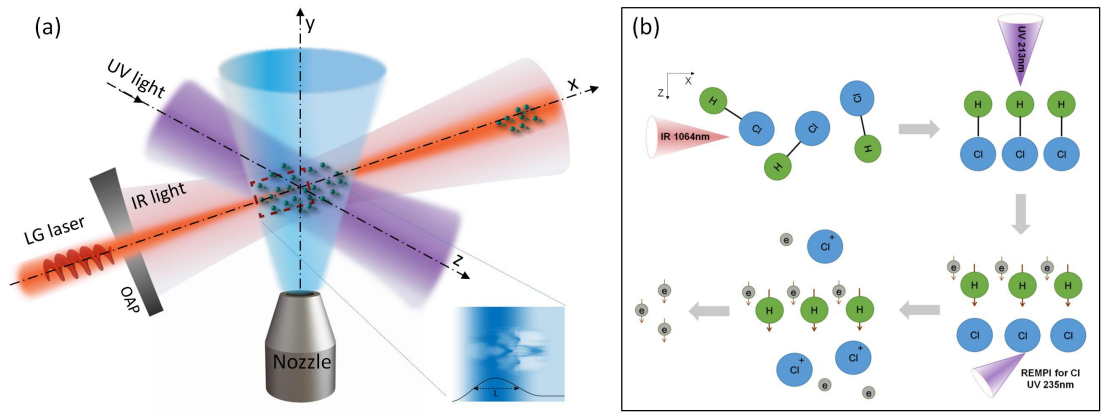

**FIG. 1** Sketch of the all-optical laser-driven polarized electron acceleration scheme with (a) the proposed experimental configuration and (b) the procedure to generate fully polarized electron target. The 1064 nm IR laser propagates along the *x* axis to align the bonds of the HCl molecules, then a UV light propagates along the *z* axis with a wavelength of 213 nm is used to photo-dissociate the HCl molecules. A 234.62 nm UV light is used to ionize the Cl atoms. Thermal expansion of the electrons creates large Coulomb field that expels the Cl ions. A fully polarized electron target is therefore produced for sequential acceleration by the LG laser pulse propagating along the *x* axis. A density ramp of length scale *L* is employed to ensure efficient electron injection.

With this in mind, the all-optical scheme is sketched in Fig.1. A spin-polarized hydrogen-atom target is prepared [16] from the photodissociation of HCl gas emitted from a nozzle. The bolds of HCl molecule are aligned by an 1064 nm infrared (IR) light (laser) first. Then double polarized hydrogen-atoms (electron and proton) are obtained via photodissociation with a circularly polarized UV light (213 nm, right-handed, perpendicular to IR) so that the spins are oriented along the UV light propagation direction. After that Cl atoms are ionized by another UV light (235 nm, not shown here) and removed by either external electric field or thermal expansion. This is followed by the driving vortex LG pulse with a well-controlled time synchronization. The latter, propagating coaxially with the

IR light, stimulates a plasma wakefield in the gas target and accelerates pre-polarized electrons to high energies.

Three relevant mechanisms may have influence on beam polarization, i.e., spin precession in electric and magnetic fields according to the Thomas-Bargmann-Michel-Telegdi(T-BMT) equation [33], the Sokolov-Ternov effect (spin flip) and the Stern-Gerlach force (gradient forces). In reality, for LWFA schemes, spin precession according to T-BMT is the main influence while others effect are negligibly small [16,17,34]. Simple estimations are given here to explain why other effects can be omitted here. Typically the polarization time for electrons via Sokolov-Ternov effect is about $T_{pol,S-T}=8m_e^5c^8/5\sqrt{3}\hbar e^5F^3\gamma_e^2$ [2,17]. In LWFA, the typical relativistic factor of electrons and field strength are $\gamma_e \sim 10^3$ and $F \sim 10^{16}$V/m, respectively. One finds $T_{pol,S-T} \sim 1\mu s$ much larger than the acceleration duration (~ns scale), thus the S-T effect can be neglected. Comparing the Stern-Gerlach force to the Lorentz force $|F_{SG}/F_L| \sim |\nabla(\bm{S}\cdot\bm{B})/\gamma_e^2 c\bm{B}m_e| \sim \hbar/\lambda_e m_e c\gamma_e^2 \ll 1$ [2,17] also suggests its minor effect in LFWA. Electron spin dynamics is therefore integrated into the PIC simulation code VLPL [35] according to the T-BMT equation [36]:

$$d\bm{s}/dt = \bm{\Omega} \times \bm{s} \qquad (1)$$

with $\bm{\Omega} = \frac{e}{m}\left(\frac{\bm{B}}{\gamma} - \frac{1}{\gamma+1}\frac{\bm{v}}{c^2}\times\bm{E}\right) + a_e\frac{e}{m}\left(\bm{B} - \frac{\gamma}{\gamma+1}\frac{\bm{v}}{c^2}(\bm{v}\cdot\bm{B}) - \frac{\bm{v}}{c^2}\times\bm{E}\right)$. Here $e$ is the fundamental charge; $m$ is the electron mass; $\bm{v}$ is the electron velocity; $\gamma=1/(1-v^2/c^2)^{-1/2}$ is the relativistic factor; $a_e=(g-2)/2 \approx 1.16\times10^{-3}$ ($g$ is the gyromagnetic factor); and the vector $\bm{s}$ is the electron spin in its rest frame [2], respectively. We have adopted the rotation matrix method in our moving particle module of the PIC code to minimize the numerical error in solving the T-BMT equation.

The LG laser propagates along the $x$ axis from left side of simulations window of 48μm(x)×48μm(y)×48μm(z) in size and 1200×600×600 in cell numbers, and passes through the fully ionized cold plasma where the electron spins are initially aligned to +z axis. We employ the $LG_{01}$ mode laser [37-39]:

$$E(r,t) = E_0 \frac{\sqrt{2}r}{w(x)} exp\left[-\frac{r^2}{w^2(x)}\right] sin^2\left(\frac{\pi t}{2\tau}\right) exp\left\{-i\left[2\,tan^{-1}\left(\frac{x}{x_R}\right) - \phi - \frac{k_0 r^2 x}{2(x^2 + x_R^2)}\right]\right\} \quad (2)$$

with $r^2=y^2+z^2$, $\phi=\arctan(z/y)$, wavelength $\lambda=800$nm, $k_0=2\pi/\lambda$, $w(z)=w_0\{[(x-x_f)^2+x_R^2]/x_R^2\}^{1/2}$, $x_R=\pi w_0^2/\lambda$ and the focusing position $x_f$, respectively. The laser pulse is linearly polarized along the $y$ axis. We also give results for a Gaussian laser for comparison . We use the density ramp injection method [40-42] with following profile: $n(x)=\{[\kappa-\Theta(\xi-L)-(\kappa-1)\Theta(\xi-2L)]\sin(\pi\xi/2L)+\Theta(\xi-L)\}n_0$, similar to the one used in [30], for better comparison. Here $\Theta(x)$ is step function, $\xi=x-x_0$, $x_0=20$μm, $L=16$μm, ratio between the peak density of the ramp and the background density $\kappa=n_p/n_0=4$, $n_p$ is the peak density while $n_0=10^{18}$cm$^{-3}$ is the background density, respectively.

## 3. Results and theoretical analysis

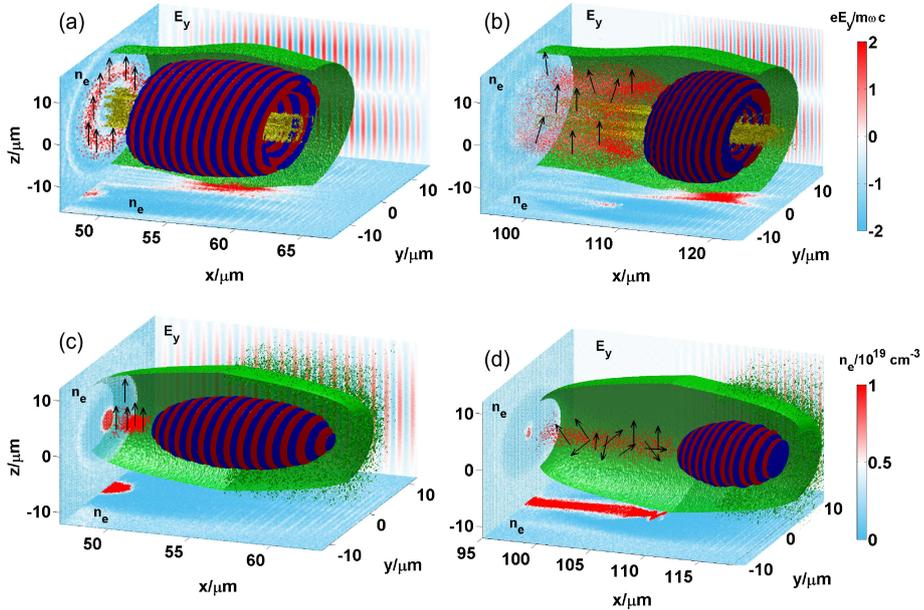

**FIG. 2** Electron density and laser field distributions (iso-surface) with $a=eE_y/m\omega c=2$ and $x_f=250\lambda$ at $75T_0$ and $150T_0$ for the LG laser (a) & (b) and for the Gaussian laser (c) & (d). The black arrows denote the electron spin directions. For the LG laser, we choose $w_0=12.5\lambda$. The Gaussian beam also has a beam radius of $12.5\lambda$. Therefore at the same laser amplitude, the pulse energy of the former is 2.7 times of that for the latter.

The acceleration processes for both laser modes are illustrated in Fig. 2, at the same peak intensity of $8.6 \times 10^{18}$ W/cm$^2$ and pulse duration of 21.4 fs. The ordinary Gaussian beam drives a bubble wakefield that traps local electrons due to the decreasing phase velocity in the plasma density bump [43]. A cylindrical electron bunch is formed in the bubble center, as depicted in Fig. 2(c) and (d). The spins of the injected electrons are well aligned in the same direction at 75$T_0$ However, at a later stage of 150$T_0$ the spin orientations are strongly diverged at different positions. Averaging the spin projections onto the $z$ axis, one finds that the overall polarization is vanishing, i.e., the electron beam is depolarized.

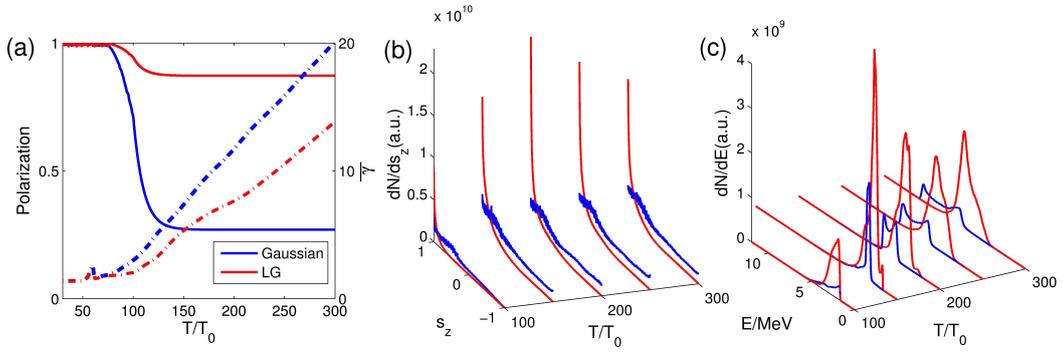

**FIG.3** Results for trapped electrons located in bubble wake. Displayed are the evolution for (a) polarization (solid lines) and electron relativistic factor $\gamma$ (averaged, dashed lines), (b) the $s_z$ distribution and (c) energy spectrum. Red and blue colors represent the LG and Gaussian laser beams.

To better understand the spin dynamics of electrons, we count all electrons injected into the bubble and calculate the polarization with $P = \sqrt{P_x^2 + P_y^2 + P_z^2}$ where $P_\alpha = \sum_i s_{\alpha i}/N$ and the averaged relativistic factor $\bar{\gamma} = \sum_i \gamma_i/N$ at each interaction time. As illustrated in Fig. 3(a), the beam polarization barely changes when interacting with the laser field (50-65$T_0$) since the initial spin directions are parallel/anti-parallel to laser magnetic field (along the +/-z axis). The polarization decreases significantly with $\gamma$ varying slowly and then followed by steady acceleration of electrons where the polarization is almost constant. The former, corresponding to the injection phase of LWFA, is when the major depolarization happens. Along the simulation, one finds that the polarization is maintained at a very high level (~88%) for the LG case, while the one for the Gaussian beam is only about 26%.

To further illustrate the polarization variation depending on the driving laser geometry, the evolution of the transverse spin component $s_z$ is shown in Fig. 3(b). During the interaction, the spins dilute towards the $s_z=-1$ end for the Gaussian pulse, while most trapped electrons accumulate at $s_z=1$ for the LG laser. From the energy spectrum in Fig. 3(c) we see that, while the cutoff and peak energies are marginally smaller for the LG case, the total number of electrons is significantly higher than that for the Gaussian beam. This generates an enormous electron-beam flux. For instance, the peak current of the LG case reaches 20 kA (polarization ~88%), about 4 times larger than in the Gaussian case (5 kA, polarization <30%). The boosted bunch charge or the peak flux benefits from the new geometry of the LG-laser-driven wakefield. The vortex beam produces a donut-shaped electron bunch in the vicinity of $r_0 \pm \triangle r/2$, as compared to a cylinder-like beam of radius $\triangle r$ for the Gaussian driver, corresponding to a cross-section area of $2\pi r_0 \triangle r$ and $\pi \triangle r^2$, respectively. For a simple estimation, we use $\triangle r \sim a^{1/2}\lambda_p(x_p)/\pi \sim 4$ μm [30,44,46] as the bunch radius of the trapped electrons and $r_0=w_0/\sqrt{2} \approx 7$ μm as the center of the trapped region for the LG case [44]. Accordingly, the peak-current ratio between the LG and the Gaussian case is about $2r_0/\triangle r \sim 3.5$, a factor that is well reproduced in simulations.

In LWFA, one usually has $B \sim B_\phi$, $E_r \sim -B_\phi$ and $v \sim v_x$ [44,46], where $B_\phi$ is the azimuthal magnetic field within the bubble. Considering $\gamma \sim 1 \gg a_e$ during the injection (see Fig.3(a)), the spin precession frequency from Eq. (1) takes the simplified form $\mathbf{\Omega} \approx eB_\phi(2+\beta_x)/2m\mathbf{e}_\phi$, To solve the equation for each particle, we separate the spin vectors $\mathbf{s}$ into the component parallel to $\mathbf{\Omega}$ with $\mathbf{s}_{//}=(\mathbf{s}\cdot\mathbf{e}_\phi)\mathbf{e}_\phi$ and perpendicular to $\mathbf{\Omega}$ with $\mathbf{s}_\perp=(\mathbf{s}\cdot\mathbf{e}_r)\mathbf{e}_r$. From the initial condition $\mathbf{s}_0=\mathbf{e}_z$, one acquires evolution of the spin vector as $\mathbf{s}=(\mathbf{e}_z\cdot\mathbf{e}_\phi)\mathbf{e}_\phi+\cos(\triangle \theta_s)(\mathbf{e}_z\cdot\mathbf{e}_r)\mathbf{e}_r+\sin(\triangle \theta_s)(\mathbf{e}_z\cdot\mathbf{e}_r)\mathbf{e}_x$. Here $\mathbf{e}_x$, $\mathbf{e}_r$ and $\mathbf{e}_\phi$ are normalized base vectors in cylindrical coordinates, the rotation angle $\triangle \theta_s \approx \langle\mathbf{\Omega}\rangle \triangle t$ depends on the time averaged precession frequency and the precession duration $\Delta t$. Then the beam polarization in each direction follows $P_x=1/N\sum\sin(\Delta\theta_s)(e_z\cdot e_r)=0$, $P_y=1/N\sum(e_z\cdot e_\phi)(e_y\cdot e_\phi)+\cos(\Delta\theta_s)(e_z\cdot e_r)(e_y\cdot e_r)=0$ and $P_z=1/N\sum(e_z\cdot e_\phi)^2+\cos(\Delta\theta_s)(e_z\cdot e_r)^2=[1+\sum\cos(\Delta\theta_s)/N]/2$, i.e., the polarization component is mainly along the $z$ direction.

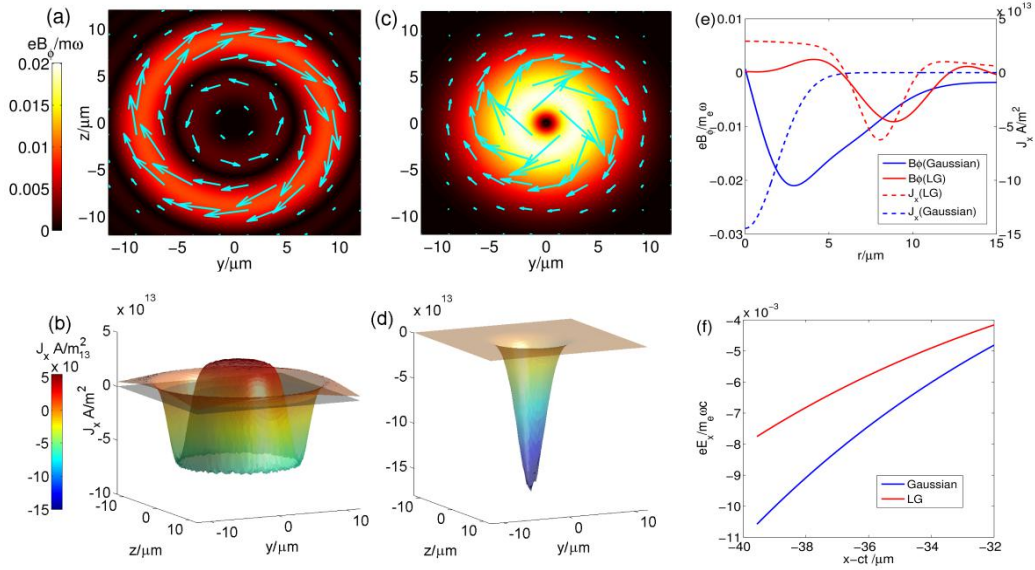

**FIG.4** Color-map of the averaged azimuthal magnetic field $B_\phi$ and current density along the $x$ direction $J_x$ during injection in the trapping region in the $y$-$z$ plane for the LG case (a) & (b) and for the Gaussian case (c) & (d). Cyan arrows show the directions and amplitudes of the averaged $B_\phi$. The averaged $B_\phi$ and $J_x$ as a function of the radius $r=\sqrt{y^2+z^2}$ (e) and the longitudinal electric field $E_x$ as a function of $x$-$ct$ (f).

The above analysis shows that the spin procession is strongly related to the azimuthal magnetic field. We averaged $B_\phi$ over the trapped electron region and display its values in the $y$-$z$ plane during injection in Fig.4. For the LG case (Fig. 4(a)), $B_\phi$ is anticlockwise in the near axis region (-5 μm < $r$ < 5 μm), decays to zero as the radius increases, and changes to clockwise in the outer region. Differently, the $B_\phi$ field for the Gaussian laser is clockwise in the whole displayed area and gradually declines (Fig. 4(c)). One sees that the peak magnetic field in the former is less than half of that in the latter. As a matter of fact, the azimuthal magnetic field $B_\phi$ in the cavity satisfies [43,44]:

$$B_\phi(r) = \frac{1}{r}\int_0^r \left(\frac{j_x}{\varepsilon_0 c^2} - \frac{\partial E_x}{c\partial x}\right) r' dr'. \qquad (3)$$

We assume $\partial E_x/\partial r \sim 0$ since in the blowout regime of LWFA the longitudinal electric field is slowly varying in the trapping area. Therefore, the field strength depends on two quantities: the longitudinal current density ($J_x$) and the electric field ($E_x$) along the $x$ axis. Since it is known for the blow-out regime that $|j_x/\varepsilon_0 c^2|/|\partial E_x/c\partial x| \sim n_p/n_0 \sim 4$ [46-48], the difference is mainly due to the self-generated magnetic field of the beam current.

For a Gaussian laser, the light intensity peaks on-axis such that trapped electrons are concentrated at the center of the bubble, leading to a well-directed current in the counter-propagation direction, as shown in Fig. 4(d). One finds the highest field strength near the symmetry axis. On the contrary, the intensity of the LG laser beam is maximized off-axis, leaving a hollow space in the propagation center. Two consequences immediately arise: First, trapped electrons are distributed in a circular ring with their density peaking at around $r = 7$ μm. The new topology significantly lowers down the electron areal density and the current density for certain amount of beam charge. Second, electrons located near the symmetry axis leak through the beam center and become the source of a counter-propagating return flux in the region of $r < 5$ μm. These effects lead to the unique current density profile for the LG driver in Fig. 4(b), where the peak current density in the trapping region is only one-third of that for a Gaussian beam. The already weakened magnetic field is further effectively compensated by the anti-clockwise field resulting from the return current. Together, they strongly reduce $B_\phi$ while maintaining the total beam charge or flux, as lined out along the radial distance in Fig. 4(e). We also note that the longitudinal accelerating field $E_x$, linearly dependent on the phase [46-48], is notably smaller in the LG case as seen from Fig. 4(f). It leads to less acceleration compared to the Gaussian case (Fig. 3(c)). A possible explanation is that the residual electron density within the cavern is larger compared to the Gaussian case due to the donut-shaped vortex distribution, which weakens $E_x$ formed by the background ions.

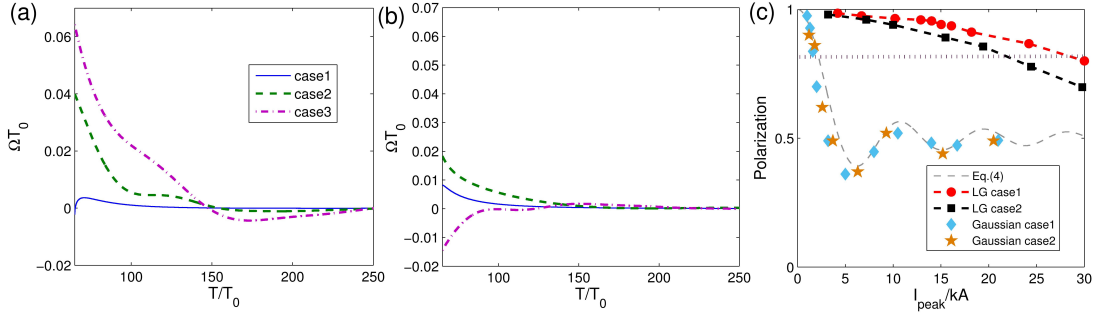

**FIG.5** The precession frequencies (normalized to $T_0^{-1}$) of (a) the Gaussian and (b) the LG case for trapped electrons at different injection radii. Blue solid, green dashed and pink dash-dotted lines represent injection positions $0, \Delta r/2, \Delta r$ for the Gaussian case and $r_0, r_0+\Delta r/2, r_0-\Delta r/2$ for the LG case respectively. (c) Beam polarization as a function of the peak current at $300T_0$ (800fs) for case 1: varying laser amplitude $a$ with fixed values of $x_f=250\lambda$, $\kappa=4$, $n_0=10^{18}\text{cm}^{-3}$; and case 2: varying $\kappa$, $n_0$ and $x_f$ at fixed laser amplitude $a=5$. The pink dotted represents polarization rate of 80%.

The essence of beam depolarization is that electron spins precess at difference frequencies. Eventually the unsynchronized spins are oriented in various directions such that the averaged spin, i.e., the beam polarization, vanishes. A direct observation of the depolarization process will be tracking the specific electrons during injection and acceleration. We chose three electrons for each case, at injection radii of $0, \Delta r/2, \Delta r$ for the Gaussian laser and $r_0, r_0+\Delta r/2, r_0-\Delta r/2$ for the LG laser, and present the precession frequencies in Fig. 5(a) and (b). In the Gaussian case, the electron spins oscillate at much higher frequencies due to larger $B_\phi$ (see Fig.4). The precession frequencies, strongly determined by the magnetic field, are diverged for different injection positions. Electrons then lose their initial spin orientations in varied paces, leading to beam depolarization. The same analysis also applies to the LG case, but the field is so much weaker that the spin evolves much slower and their directions for off-axis electrons remain well-aligned during the whole interaction process.

Considering $\langle\beta_x\rangle\sim 1/2$ during the injection phase, the average precesion frequencies can be written as $\langle\Omega\rangle\approx 5e\langle B_\phi\rangle/4m$. As we noted from Fig.5(a) and 5(b), the electrons with smaller injected radius seems trapped faster($\Omega$ declines faster). Taking this into account, we consider that electrons with injection radius $r_i$ undergo $2\pi|r_i\text{-}r_0|$ for LG case and $\pi r_i$ for gaussian case during injection phase. With this in mind, we treat the injection time for $\Delta t\sim\pi|r_i\text{-}r_0|/<\beta_x>\sim 4\pi|r_i\text{-}r_0|/c$ LG case and $\Delta t\sim 2\pi r_i/c$ Gaussian case, which is close to previous studies where considering trapped electrons as a whole with $\Delta t\approx 4a^{1/2}\lambda_p/\pi c$ [30,49,50]. Assuming the injection density is homogenous among whole injection region, the whole polarization after injection is given as $P = [1+\sum \cos(\Delta\theta_s)/N]/2 \approx 0.5 + \left[\int_{r_0-\Delta r/2}^{r_0+\Delta r/2} \cos\left(5\pi e<B_\phi(r)>\frac{r-r_0}{mc}\right)rdr\right]/(2r_0\Delta r)$ for the LG case and $P \approx 0.5 + \left[\int_0^{\Delta r} \cos\left(5\pi e<B_\phi(r)>\frac{r}{2mc}\right)rdr\right]/\Delta r^2$ for the Gaussian case. We calculate a polarization of about 0.91 for the LG case and 0.31 for the Gaussian case from the $B_\phi$ profiles in Fig. 4(e), in good agreement with the simulation results from Fig. 3(a).

After electrons gaining sufficiently high energies the spin directions remain almost unchanged (see Fig.3(a)). In fact, the spin precession angle $\Delta\theta_s$ can be roughly estimated during this phase. In steady acceleration, one has $E_r\sim cB_\phi, v\sim v_x=\beta_x c$ [41] (see also Fig.4), therefore the precession frequency can be written as $\mathbf{\Omega}\approx eB_\phi/m\gamma(\gamma+1)$, for $\beta_x\approx 1$ and $\gamma\gg 1$, suggesting the electron spin precession is slowed down due to $\gamma\gg 1$. Substituting the equation of motion for electrons $\gamma/\Delta t\sim d\gamma/dt\sim eE_x/mc$ into the precession frequency, we finally obtain $\Delta\theta_s\sim cB_\phi/E_x(\gamma+1)\sim 1/\gamma\ll 1$, i.e., the spin change is negligible during this phase.

Spin depolarization imposes strong restrictions on the charge or the current of the electron beam from LWFA. One can find out the criteria for the Gaussian laser beam by taking $B_\phi\sim B_0 r/\Delta r\sim en_p r/8\varepsilon_0 c$ [51, 52] and the radius of the injection volume $\Delta r=(I_{peak}/\pi en_p c)^{1/2}$ with injection time $\Delta t=2\pi r/c$. The polarization is obtained from the statistic average of spins in the injection volume:

$$P = \frac{1+\sum\cos(\Delta\theta_s)/N}{2} = \frac{1+\text{sinc}(\alpha I_{peak})}{2} \qquad (4)$$

where sinc($x$)=sin($x$)/$x$ and $\alpha=5\pi e/16m\varepsilon_0 c^3$. Hence to retain polarization >80% the criteria $\alpha I_{peak}$<1.8 applies, corresponding to the restriction $I_{peak}$ <2.5kA. In Fig.5(c) we show systematic scans at various laser amplitudes $a$, background densities $n_0$, density ratio $\kappa$ and focal position $x_f$. In all cases the beam current is limited to $I_{peak}$ <2.2kA for the Gaussian laser to preserve polarizations over 80% (pink dotted line in Fig.5(c)), consistent with the prediction in Eq. (4). However, the limitation on the beam flux is released because of the vortex beam structure. The peak current for the LG case $I_{peak}$ reaches ~20kA where the polarization is ~90%, an order of magnitudes higher.

We find interesting oscillations of the beam polarization for the Gaussian laser in Fig. 5(c). This can be seen from Eq.(4), where the second term in the numerator periodically changes with the peak current, describing the oscillation very well. It is known that only the component perpendicular to the magnetic field in the rest frame of the electron $s_\perp$ processes. For larger peak current, the precession frequencies increase such that the transverse beam polarization, averaged over accelerated electrons, vanishes for $\Delta\theta_{sm}$>>2$\pi$. In other words, $P_\perp=\sum\cos(\Delta\theta_s)/N \approx 0$, leaving the non-changing parallel to the magnetic field in the rest frame of the electron. Therefore the oscillation converges to $P=P_{//}$=1/2 for growing peak current.

Our simulations are carried out for initial electron polarization along the laser magnetic field. Nevertheless, our scenario is valid for all starting spin directions. For initial polarizations $s_0=e_x$ (parallel to propagation direction) and $s_0=e_y$ (parallel to electric direction) the results are illustrated in Fig.6(a). The polarizations of both cases are 81.2% for $s_0=e_x$ and 88.6% for $s_0=e_y$, validating the novel effect of the vortex laser geometry. For the longitudinal pre-polarized case ($s_0=e_x$), the polarization oscillation still happens as shown in Fig. 6(b). Unlike in the transverse polarization case, one has $P=P_x=\sum\cos(\Delta\theta_s)/N$ for $s_0=e_x$, converging to 0 polarization at sufficiently large currents. The restriction on peak current is also stronger, e.g. to preserve polarization >80%, $I_{peak}$ <1.5kA.

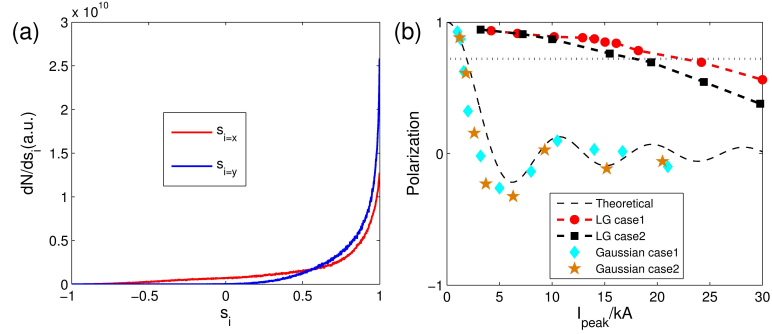

**Fig.6** (a)The Spin component distribution at $300T_0$($T_0=\lambda/c$=2.67fs is the laser cycle) for LG. The red lines denote $s_x$ distribution for $s_x$=1 at initially (parallel to propagation direction) while blue lines for $s_y$ distribution for $s_y$=1 at initially(perpendicular to propagation direction). (b) Beam polarization as a function of the peak current at $300T_0$ (800fs) for $s_0=e_x$ with the same case in Fig.5(c).

## 4. Conclusions

In conclusion, we proposed a promising scheme to generate polarized electron beams via LWFA driven by a vortex Laguerre-Gaussian Laser. According to our 3D-PIC simulations involving particle spin dynamics, electron beam is of polarization over 80% is achieved with high beam charge and peak flux. Compared to the Gaussian laser driven acceleration, the restriction on the electron beam current density is released, thanks to the novel topology of the vortex LG laser. The scheme relies on an all optical set-up that is accessible at state-of-the-art facilities.

## Acknowledgements

We would like to thank Prof. Andreas Lehrach for helpful discussions. This work is supported by the Strategic Priority Research Program of Chinese Academy of Sciences (Grant No. XDB 16010000), the National Science Foundation of China (No. 11875307) and the Recruitment Program for Young Professionals. The German authors acknowledge support through the HGF-ATHENA project.